# DARK AND BRIGHT PATTERNS IN COOKIE CONSENT REQUESTS

Paul Graßl[1], Hanna Schraffenberger[1], Frederik Zuiderveen Borgesius[1,2] and Moniek Buijzen[3,4]

## Abstract

Dark patterns are (evil) design nudges that steer people's behaviour through persuasive interface design. Increasingly found in cookie consent requests, they possibly undermine principles of EU privacy law. In two preregistered online experiments we investigated the effects of three common design nudges (default, aesthetic manipulation, obstruction) on users' consent decisions and their perception of control over their personal data in these situations. In the first experiment ($N$ = 228) we explored the effects of design nudges towards the privacy-unfriendly option (dark patterns). The experiment revealed that most participants agreed to all consent requests regardless of dark design nudges. Unexpectedly, despite generally low levels of perceived control, obstructing the privacy-friendly option led to more rather than less perceived control. In the second experiment ($N$ = 255) we reversed the direction of the design nudges towards the privacy-friendly option, which we title "bright patterns". This time the obstruction and default nudges swayed people effectively towards the privacy-friendly option, while the result regarding perceived control stayed the same compared to Experiment 1. Overall, our findings suggest that many current implementations of cookie consent requests do not enable meaningful choices by internet users, and are thus not in line with the intention of the EU policymakers. We also explore how policymakers could address the problem.

1 iHub, Radboud University, Nijmegen, The Netherlands
2 Institute for Computing and Information Sciences (iCIS), Radboud University, Nijmegen, The Netherlands
3 Behavioural Science Institute, Radboud University, Nijmegen, The Netherlands
4 Erasmus School of Social and Behavioural Sciences, Erasmus University Rotterdam, Rotterdam, The Netherlands

## 1  INTRODUCTION

Whenever people are browsing the web they face privacy decisions in the form of cookie consent requests. The goal of cookie consent requests (under the EU's ePrivacy Directive) is a) to inform users about the goal of the cookies, and b) ask users for their consent. To give online users control over their personal data, the ePrivacy Directive only allows the use of tracking cookies (and similar tracking technologies) after the user has given his or her prior consent.

To ensure that users understand the decision they make with a consent request, consent (for tracking cookies) in the ePrivacy Directive must be interpreted in line with the strict criteria for valid consent in the General Data Protection Regulation (GDPR 2016); we refer to the two legal acts together as "EU privacy law". These criteria include that valid "consent" of the internet user (data subject) requires a "freely given, specific, informed and unambiguous indication of the data subject's wishes by which he or she, by a statement or by a clear affirmative action, signifies agreement" (GDPR, 2016, article 4(1)). In that, EU law appears to assume that people make deliberate and well-informed privacy choices. This assumption corresponds to a prominent model of privacy decision making, the privacy calculus theory, which presumes people's behaviour to be fundamentally rational and privacy decisions to be made through conscious weighing of the costs and benefits of each choice option (Laufer & Wolfe, 1977). But do people in practice perceive control over their personal data and show deliberative rational decision behaviour in the context of cookie consent requests?

This is questionable in light of a new trend of using "dark patterns" in cookie consent requests, which aim to influence users' privacy decisions (e.g., through pre-ticked boxes or highlighted options; Forbrukerrådet, 2018). Dark patterns are (evil) design nudges, which steer users against their best interest towards a certain choice through persuasive interface design (Brignull, n.d.; Gray, Kou, Battles, Hoggatt, & Toombs, 2018). Originally, nudging means influencing the decisions of individuals or groups towards good choices (as judged by themselves) through minor changes in the choice environment without compromising freedom of choice (a prominent example is a fly painted on a urinal in a public men's toilet to prevent urine spillage; Thaler & Sunstein, 2009).

The use of dark patterns can be problematic for legal as well as ethical reasons. While the GDPR (2016) does not explicitly ban all dark patterns, they do breach the spirit of the GDPR. Ethically, dark patterns (and nudges in general) may lead users to make choices that are not in their interest and deprive users of their control (Forbrukerrådet, 2018; Schubert, 2015). In fact, if a nudge is used for evil, Thaler (2018) refuses to call it "nudge", but rather "sludge". His colleague Sunstein (2016b) states two conditions to assess whether a manipulation is ethically objectionable: (1) when the goals of the manipulator are self-interested and (2) when the manipulation subverts the chooser's deliberative capacities. Dark patterns meet the first condition because they are used in the interest of the manipulator to collect personal data. The second criterion, as we will argue in the next paragraphs, is met as well because dark patterns push users to make quick heuristic decisions rather than slow and deliberate ones.

EU privacy law and the privacy calculus theory assume that people make privacy decisions with what Kahneman (2011) calls System 2, that is the slow and consciously reasoning part of us. However, considering evidence from a multi-disciplinary literature assessment from Acquisti et al. (2017), it cannot be assumed that people behave purely rational in privacy decision situations. Rather, people apply heuristics - mental shortcuts in decision-making - and fall back to cognitive or behavioural biases, which work on the quick, heuristic System 1 (Sunstein, 2016a).

Cookie consent requests feature several characteristics that make people prone to applying heuristics. First, there is an information asymmetry between the user confronted with the consent request and the company asking for it. The user has access to less information regarding the purpose of data collection and possible future usage of it than the data controller. Second, consent requests often use ambiguous language (e.g., the data may be used for a certain cause) creating a decision under uncertainty for the user because not all possible outcomes are known. Acquisti et al. (2017) argue that these circumstances facilitate the application of heuristics, given that human rationality is limited to the available cognitive resources and the available time (based on the concept of bounded rationality; Simon, 1957). Third, people's privacy decisions are influenced by several cognitive biases, such as the status-quo-bias (individuals' preference for default choices) or the salience-bias (individuals' tendency to focus on prominent features). These three circumstances of cookie consent requests likely facilitate the mechanism of dark patterns, which targets mainly the intuitive, heuristic System 1 (Bösch, Erb, Kargl, Kopp, & Pfattheicher, 2016).

While there are many examples of the use of dark patterns in practice (see Brignull, n.d.; Fansher, Chivukula, & Gray, 2018; Forbrukerrådet, 2018), the field of privacy and data protection lacks research in this regard. The few studies that focused on the effects of dark patterns were conducted either with a non-representative sample (e.g., only students or young university-educated people; Machuletz & Böhme, 2019; Nouwens, Liccardi, Veale, Karger, & Kagal, 2020) or in a context that cannot be generalised easily (e.g., participants were told to have been automatically signed up for a costly identity-theft protection service; Luguri & Strahilevitz, 2019). Solely Utz, Degeling, Fahl, Schaub, and Holz (2019) demonstrated adequately that the use of dark patterns possibly influences a user's consent decisions, however, giving no clear answer on how to deal with the underlying problem of an overwhelming number of consent requests which may lead to indifference towards them over time.

Therefore, it is crucial to gain a better understanding of the effects of design nudges in cookie consent requests and to assess whether a) the understanding of privacy decision making in EU privacy law represents reality, and b) whether users perceive control over their personal data through consent requests. We investigated these aims in two online experiments: Experiment 1 focused on the effects of dark patterns on people's consent decisions and their perception of control over their personal data. In a follow-up experiment (Experiment 2), we reversed the direction of the design nudges (i.e., towards the privacy-friendly option) to see how this affects people's consent behaviour and their perceived control compared to the first experiment (we titled such privacy-friendly design nudges "bright patterns"). Following, we will briefly introduce and outline the two experiments. After that, we focus in more detail first on Experiment 1 and then on Experiment 2. We end with a general discussion.

### 1.1 Experiment 1: Dark patterns in cookie consent requests

In our first experiment, the research questions were: given a cookie consent request with two choice options (privacy-friendly vs. privacy-unfriendly), do dark patterns lead users to choose the privacy-unfriendly option more often than the privacy-friendly option, even if the privacy-friendly option is rationally superior? And do dark patterns deprive users of their perceived control over their personal data? Specifically, we focused on the effects of three of the most common dark patterns, that is (1) default, (2) aesthetic manipulation and (3) obstruction (Fansher et al., 2018).

Default refers to any situation where one option is preselected prior to any action of the user, for example when the option to agree to a privacy policy is selected by default (Gray et al., 2018). Aesthetic manipulation refers to the act of giving "one option visual or interactive precedence over others", for example when one out of two choice buttons is coloured blue while the other one is simply grey (also called "false hierarchy"; Gray et al., 2018, p. 7). Obstruction means making an interaction more effortful than it needs to be to dissuade the user from a certain action or choice, for example when the option to opt out of online tracking is not presented together with the opt-in option but can only be reached by clicking through several submenus.

Following this design nudge (towards choosing the privacy-unfriendly option) can be considered a non-rational choice if the privacy-friendly option has more benefits (i.e., is rationally superior) than the privacy-unfriendly option (Archer, 2013). In Experiment 1, we presented the privacy-unfriendly option (i.e., allowing web tracking) in such a way that choosing this option could lead to losing control over one's personal data without providing any benefit (such as more relevant advertising). Hence privacy calculus theory would predict that people choose the privacy-friendly option (Smith, Dinev, & Xu, 2011). Deviations from this prediction indicate that people engage in privacy decisions (in the context of cookie consent requests) with the automatic, heuristic System 1, rather than with the rational, deliberate System 2.

We formulated the following hypotheses. In a cookie consent request situation with two choice options (privacy-friendly vs. privacy-unfriendly), where the privacy-friendly option is rationally superior,

**Hypotheses 1a/b/c**: participants will be more likely to choose the privacy-unfriendly option (compared to privacy-friendly) when the privacy-unfriendly option is (H1a) preselected, (H1b) visually more salient or (H1c) the alternative (privacy-friendly) option is obstructed.

**Hypotheses 2a/b/c**: participants report lower levels of perceived control over their personal data when the privacy-unfriendly option is (H2a) preselected, (H2b) visually more salient or (H2c) the alternative (privacy-friendly) option is obstructed.

Because little is known about the effects of dark patterns in cookie consent requests, the first study focused on their main effects rather than possible (and more speculative) interaction or moderation effects, in order to create a solid basis for further investigation. Nevertheless, we repeatedly highlighted that deliberating about a decision indicates System 2 behaviour. Little conscious deliberation, on the other hand, is associated with heuristic System 1 decision making (Albar & Jetter, 2009), which dark patterns seem to target. Therefore, we explored the possible moderating role of deliberation in the decision process. We hypothesised that more deliberation would reduce the effects of the dark patterns on the consent decisions and on the level of control that people perceive.

## 1.2 Experiment 2: Bright patterns in cookie consent requests

In the follow-up experiment, we reversed the direction of the design nudges (i.e., towards the privacy-friendly option) to see how this affects people's consent decisions and their perception of control over their personal data. We formulated the follow-up research questions based on the results from Experiment 1, where most people agreed to all consent requests in a default manner. The two research questions were thus: given a cookie consent request situation with two options (privacy-friendly vs. privacy-unfriendly), do bright patterns lead users to choose the privacy-friendly option more often than the privacy-unfriendly option (despite the previously observed default behaviour towards the privacy-unfriendly option)? And do bright patterns deprive users of their perceived control over their personal data in a similar way as dark patterns (given that any form of System 1 nudge compromises one's perception of control to some extent; Schubert, 2015; Sunstein, 2016a)?

We hypothesised that in a cookie consent request situation with two choice options (privacy-friendly vs. privacy-unfriendly),

**Hypotheses 3a/b/c**: participants will be more likely to choose the privacy-friendly option (compared to privacy-unfriendly) when the privacy-friendly option is (H3a) preselected, (H3b) visually more salient or (H3c) the alternative (privacy-unfriendly) option is obstructed.

**Hypotheses 4a/b/c**: participants report lower levels of perceived control over their personal data when the privacy-friendly option is (H4a) preselected, (H4b) visually more salient or (H4c) the alternative (privacy-unfriendly) option is obstructed.

In addition to the design nudges, other factors may influence whether a person acts in a rather fast and heuristic or more deliberate manner on privacy decisions. Based on evidence from previous research (Awad & Krishnan, 2006; Lai & Hui, 2006; Malhotra, Kim, & Agarwal, 2004) we controlled for general privacy concerns in both experiments. Additionally, we investigated in Experiment 2 whether controlling for privacy fatigue, as proposed by Choi, Park, and Jung (2018), instead of privacy concerns leads to different results.

## 2 EXPERIMENT 1

### 2.1 Method

Before running Experiment 1, we preregistered our sample size estimation, hypotheses and statistical analysis. The preregistration, the code of the study application, all used materials, data, and analysis scripts are available on the Open Science Framework (https://osf.io/c7qza/). Information about the used R version and all packages can be found in Appendix A.

#### 2.1.1 Procedure and Design

The online experiment followed a within-subjects design where participants were asked to review eight news websites (shown in random order) and report on their first impression of the visual design of each news website. We used this cover story to create a realistic setting for the presentation of cookie consent requests and disguise the true purpose of the study. Each news website displayed an overlaying cookie consent request when being visited (while the rest of the website was dimmed at first), offering two choice possibilities: allow the website and other third parties to collect data and to track user's web behaviour (privacy-unfriendly), versus not allowing such data collection and web tracking (privacy-friendly).

After the participant made a choice, the overlaying consent request disappeared and the news website was shown (no matter which option the participant had selected), but only for three seconds to fit the cover story about first impressions. Regardless of the participants' choice, we did not track their behaviour nor collected more data than that necessary for the experiment (i.e., we only recorded the consent decision). Each news website visit was followed by three questions about the participant's first impression of the design of the news website (for the sake of the cover story). After reviewing all news websites (which corresponds to part 1 of the experiment), we presented the eight consent requests again (one by one in the form of screenshots), and asked participants how much control they felt each consent request gave them over their personal data and how much they had deliberated on their decision. Additionally, for each consent request (presented as a screenshot), we asked manipulation check questions about whether participants had read the consent information and could recall the option they had chosen. Lastly, we assessed each participant's general privacy concerns and asked control questions about individual browser setup and device type. At the end of the study, we debriefed participants about the cover story and the true purpose of the experiment.

### *2.1.2    Web application and Materials*

#### 2.1.2.1    Web application

To run our online experiment, we set up a web application using the Python framework Flask (Lord, Mönnich, Ronacher, & Unterwaditzer, 2010). The application was hosted on a university server. We conducted a preliminary pilot study to test the credibility of our cover story. Four bachelor students were asked to do the study while thinking aloud, showing that the cover story worked as intended. We used eight different news website templates, which are licensed under the Creative Commons Attribution 3.0 (Colorbib, 2019). The news websites were called Avision, Megazine, Motivemag, Quitelight, Techmag, Technews, Viral and Webmag. We adjusted the templates partly in functionality (e.g., hyperlinks were disabled), content (e.g., exchange placeholder text such as "lorem ipsum" with plausible news content) and design to fit the purpose of our study. To achieve additionally required functionality for the online experiment, such as building multi-step consent requests (i.e., obstruction manipulation) or detecting when participants clicked on the back button, we used code solutions from An (2019) and Brooke (2011), respectively, which are available under the MIT license. Two examples of the used websites can be found in Appendix B, while the rest can be found on the Open Science Framework.

#### 2.1.2.2    Consent requests

For each news website, we created a cookie consent request, which appeared as an overlay when a participant was directed to the news website. The general layout and text of the consent requests were inspired by a corpus consisting of consent requests of several popular news websites and big tech companies (corpus available on the Open Science Framework). The aim was to create cookie consent requests that resemble many of such consent requests used in practice. Whereas we kept the main characteristics (e.g., the content of the provided text) of the consent requests constant across all conditions, we changed minor design details (e.g., font type, order of the sentences in the text, colour of the consent box edges etc.) of each consent request, to make them look slightly different from each other and to support the cover story about eight independent, external news websites. To have an indication of non-rational behaviour, the consent requests provided no information about any benefit of choosing the privacy-unfriendly option "Agree" (e.g., better-targeted advertising), which only left the cost of potentially losing control over ones' personal data when agreeing to the policy (i.e., allowing web tracking). Hence, choosing the privacy-unfriendly option "Agree" can be considered a non-rational choice (an example consent request text can be found in Appendix C).

Whereas the general layout of the consent requests was consistent, each request contained one out of eight possible combinations of the three dark patterns (1) default, (2) aesthetic manipulation and (3) obstruction. The statistical model we used (mixed-effects model) required the inclusion of all possible combinations of the independent variables (i.e., the dark patterns) to accurately estimate the effect of each predictor. Default was represented by a preselected "Agree" radio button on the websites Quitelight, Techmag, Technews and Webmag (Figure 1 shows one example consent request; screenshots of all consent requests can be found on the Open Science Framework). Aesthetic manipulation was represented by a blue coloured "Agree" button on the websites Megazine, Techmag, Viral and Webmag. Obstruction was represented by the option "Manage options" instead of "Do Not Agree" on the websites Motivemag, Technews, Viral and Webmag. Participants could only choose "Do Not Agree" after selecting "Manage options". The consent request of the website Avision represented the baseline condition with none of the three design nudges included (see figure in Appendix B1).

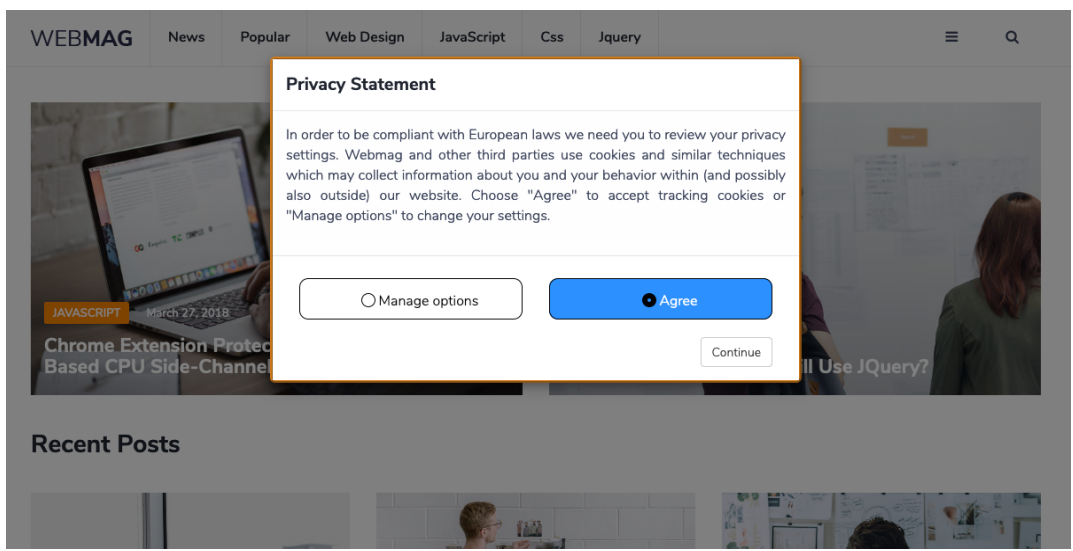

*Figure 1. Example consent request featuring all three dark patterns default, aesthetic manipulation and obstruction. Website: Webmag*

2.1.2.3  Measures

For each consent request, we recorded a participant's consent decision and assessed his or her level of perceived control, level of deliberation and several control questions regarding his or her attentiveness during the decision process. Further, we asked participants to report on their general privacy concerns and personal browser setup.

To measure how much control participants felt each consent request gave them over their personal data we built on the Perceived Control scale from Xu (2007). We adjusted the formulation of the items to fit the purpose of the study (see Table 1). Participants could indicate their perceived level of control over their personal data on a slider ranging from *Not at all* to *Complete* (higher values indicate more perceived control). We used the average of all five items as the final outcome variable perceived control in the statistical analysis (range: 0 - 100, $M$ = 31.80, $SD$ = 28.54). Further, the perceived control measure showed very good internal consistency with a raw Cronbach's $\alpha$ = 0.99 (none of the individual items increased the overall $\alpha$ if being dropped).

We assessed how much participants deliberated about their decision by asking "How much did you think about your decision before clicking on one option?" (formulation of the item was adapted for the present study; Dijksterhuis, Bos, Nordgren, & van Baaren, 2006). Participants could indicate the level of deliberation on a slider ranging from *Not at all* to *A great deal* (range: 0 - 100, $M$ = 20.99, $SD$ = 25.33). Lastly, we used the Global Information Privacy Concern scale from Malhotra et al. (2004) to assess general privacy concerns (on a seven-point scale ranging from *Strongly disagree* to *Strongly agree*, range: 1 - 7, $M$ = 4.13, $SD$ = 1.24). For the statistical analysis, we used the average score of the three items, which formed the scale. The measure General Privacy Concerns showed good internal consistency with a raw Cronbach's $\alpha$ = 0.79 (again none of the individual items increased the overall $\alpha$ if being dropped).

We included several manipulation checks and control questions to get a better understanding of the participants' behaviour during the study. When reviewing each consent request (in the form of a screenshot), we asked whether the participant had read the consent information (in 10.1% of the cases "Read it completely", 49.6% "Skimmed it", 40.3% "Did not read it at all") before clicking on an option and whether they remembered which option ("Agree", "Do Not Agree") they had chosen (2.6% of all consent decisions could not be remembered correctly). Further, we asked whether participants had installed a browser plugin, which handles or deletes cookies (31.1% "Yes", 68.9% "No").

Table 1. Perceived control questionnaire items

| Number | Question |
| --- | --- |
| 1 | How much control did you feel the consent form gave you over the amount of your personal information collected by the company? |
| 2 | How much control did you feel the consent form gave you over who can get access to your personal information? |
| 3 | How much control did you feel the consent form gave you over your personal information that has been released? |
| 4 | How much control did you feel the consent form gave you over how your personal information is being used by the company? |
| 5 | Overall, how much did the consent form made you feel in control over your personal information provided to the company? |

*Note*. $M$ = 31.80, $SD$ = 28.54, range: 0 - 100, Cronbach's $\alpha$ = 0.99 (raw)

### 2.1.3 Participants

We recruited a total of $N$ = 228 participants for Experiment 1 via the crowdsourcing platform Prolific Academic. This sample size was initially determined for a frequentist regression analysis as preregistered for Experiment 1 (detailed information about the power estimation can be found via the Open Science Framework link provided above).

Inclusion criteria for study participation were an age between 18 and 65 years (to represent a broad range of society) and a current living location in the United Kingdom (to minimise noise in the data because of cultural differences we restricted the study to the biggest participant pool within Prolific Academic). Participants were compensated with 1.70GBP for the successful completion of the study, which was estimated to take around 12 minutes (8.50GBP/h). On average it took participants 9.79 minutes ($SD$ = 4.02) to complete the study. We left 33 participants out of this calculation because they showed very long completion times, indicating that they divided the study over several days. Yet, their consent behaviour did not seem to differ from the rest of the sample and thus they were kept for analysis. Additionally, we found that only 5 participants had completed the experiment in less than 5 minutes (but not under 3 minutes). Because of that low number, we kept them in the sample. We excluded participants who could not finish the study due to technical problems.

The total sample population consisted of 137 females (60.1%), 91 males (39.9%) and had a mean age of 36.02 years ($SD$ = 11.62). Of all 228 participants who took part in the experiment, 35 dropped out in the second part of the study (i.e., after reviewing the eight news websites). Because none of the dropouts happened during the completion of a questionnaire (only in between) and no prevalent pattern of missingness was detected (e.g., the consent behaviour did not differ between participants with complete cases and those who would drop out later on), we found all participants' data eligible for analysis.

### 2.1.4   Data Analyses

As mentioned earlier, we initially conducted a frequentist regression analysis for Experiment 1. However, we decided later to use a Bayesian framework for Experiment 2 for two reasons: Firstly, Bayesian model results fit better with how people think about and interpret parameter estimates compared to frequentist models (Morey, Hoekstra, Rouder, Lee, & Wagenmakers, 2016). Secondly, Bayesian regression models turn out often to be superior to frequentist models when it comes to multilevel structured data (Browne & Draper, 2006; Bryan & Jenkins, 2016). Therefore, we reran the analysis of Experiment 1 for consistency purposes using Bayesian modelling (the pattern of results did not differ between the frequentist and the Bayesian approach). All reported statistics refer to the Bayesian models.

Instead of classic significance testing, we used 95% credible intervals (CrI) to decide whether a given parameter has a substantial impact on the outcome. Credible intervals indicate a range within which the parameter of interest lies with a probability of X% (we used 95%), given the data. If the credible interval of a parameter does not include zero (zero would mean no effect) we assume a substantial effect of the corresponding variable on the outcome. Credible intervals are different from frequentist confidence intervals, however, the latter gets often incorrectly interpreted as the former (Morey et al., 2016). The analysis was conducted using Stan (Carpenter et al., 2017) called via the package brms (Bürkner, 2017) within the R environment (R Core Team, 2020).

For each of the two dependent variables (consent decision and level of perceived control), we fit separate models with a maximal random-effects structure, following the advice of Barr, Levy, Scheepers, and Tily (2013). Thus, each main model included a per-participant random adjustment to the fixed intercept and a per-participant random adjustment to the slope of each within-subject variable (default, aesthetic manipulation and obstruction). Further, main models included general privacy concerns as a control variable.

To fit exploratory models, we added deliberation as a moderator to the aforementioned design of the main models. Specifically, deliberation was present as a fixed effect and part of an interaction with each of the three main predictor variables. Additionally, exploratory models added a per-participant random adjustment to the slope of the main effect of deliberation and each interaction term with it.

We used the *Bernoulli* distribution as the model family for all models with consent decision as the dependent variable. The estimated models had thus the form of:

$$\begin{aligned}
y_i &\sim Bernoulli(p_i) \\
logit(p_i) &= \alpha_{j[i]} + \beta_{j[i]} x_i \\
\alpha_{j[i]} &\sim Normal(\alpha, \sigma_\alpha) \\
\beta_{j[i]} &\sim Normal(\beta, \sigma_\beta) \\
\alpha &\sim Normal(0, 10) \\
\beta &\sim Normal(0, 5) \\
\sigma_\alpha &\sim Cauchy(0, 2.5) \\
\sigma_\beta &\sim Cauchy(0, 2.5)
\end{aligned}$$

In this mixed-effects model, $i$ refers to each element of $y$ (i.e., the observed consent decisions), and $j$ denotes the grouping factor, the participant. For models with perceived control as the dependent variable, we chose the *Beta* distribution as the model family to mirror the continuous but interval restricted nature (0,1) of the outcome best (following Ferrari & Cribari-Neto, 2004). We estimated the models in the following manner:

$$\begin{aligned}
y_i &\sim Beta(\mu_i, \phi) \\
logit(\mu_i) &= \alpha_{j[i]} + \beta_{j[i]} x_i \\
\alpha_{j[i]} &\sim Normal(\alpha, \sigma_\alpha) \\
\beta_{j[i]} &\sim Normal(\beta, \sigma_\beta) \\
\alpha &\sim Normal(0, 10) \\
\beta &\sim Normal(0, 5) \\
\sigma_\alpha &\sim Cauchy(0, 2.5) \\
\sigma_\beta &\sim Cauchy(0, 2.5) \\
\phi &\sim Gamma(0.01, 0.01)
\end{aligned}$$

Due to a lack of previous literature to build on in terms of expected effect sizes, we applied only weakly informative priors on parameter estimates in all models.

## 2.2 Results

### 2.2.1 Main Analyses

To investigate our first set of hypotheses 1a/b/c (stating that dark patterns will sway people towards the "Agree" option) we first visualise the recorded consent decisions for each news website (see Figure 2). We observed that in the majority of cases (93.8%) people chose to agree to the consent requests. Moreover, most people chose always the same consent option for each news website, suggesting that nudging did not seem to matter for their decision (only 4.0% of all participants changed their consent behaviour between conditions).

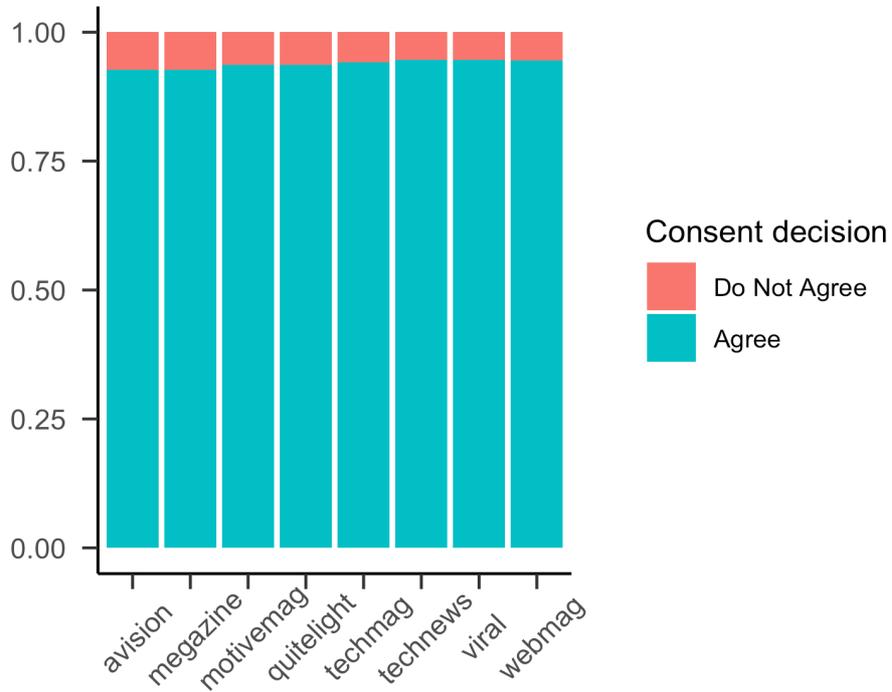

*Figure 3. Consent decisions (proportional) by condition (different news websites)*

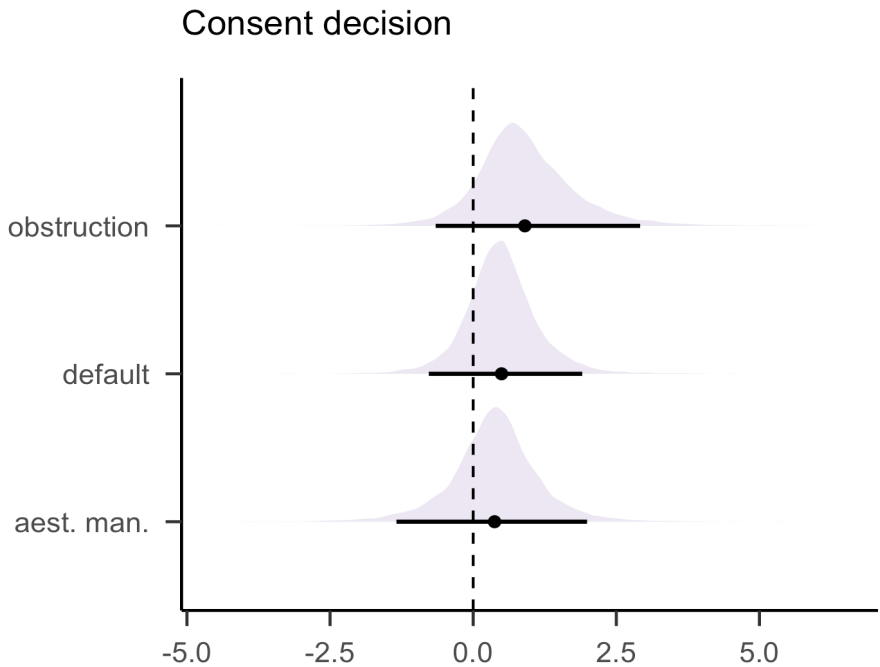

*Figure 2. Posterior distributions with mean and 95% credible interval for the predictors obstruction, default and aesthetic manipulation (outcome consent decision)*

Our results confirm this suggestion as we did not find support for our hypotheses H1a, H1b and H1c, meaning that there was no substantial effect of default, $\beta$ = 0.50 (0.66), CrI 95% [-0.78, 1.90], OR = 1.64, aesthetic manipulation, $\beta$ = 0.37 (0.80), CrI 95% [-1.34, 1.99], OR = 1.45, or obstruction, $\beta$ = 0.90 (0.87), CrI 95% [-0.66, 2.92], OR = 2.47, on the outcome consent decision (see Figure 3). The pattern of results did not change when additionally accounting for the previous consent decision of a participant (although this was a good predictor of each consent decision given that most people did not vary their consent behaviour between conditions) or whether a participant had a browser plugin installed that handles or deletes cookies. Regarding our second set of hypotheses 2a/b/c, we did not find that the dark patterns made people perceive less control over their personal data. To our surprise, however, we found the design nudge obstruction to have the opposite effect: people reported more rather than less perceived control over their personal data when the "Do Not Agree" option was obstructed by "Manage options".

More specifically, obstruction showed a small positive effect, $\beta$ = 0.11 (0.03), CrI 95% [0.05, 0.17], OR = 1.11 (see Figure 4). Hence H2c was not supported. Further, we did not find support for hypotheses H2a and H2b concerning the effects of default, $\beta$ = 0.01 (0.01), CrI 95% [-0.02, 0.04], OR = 1.01, and aesthetic manipulation, $\beta$ = 0.01 (0.02), CrI 95% [-0.03, 0.05], OR = 1.01.

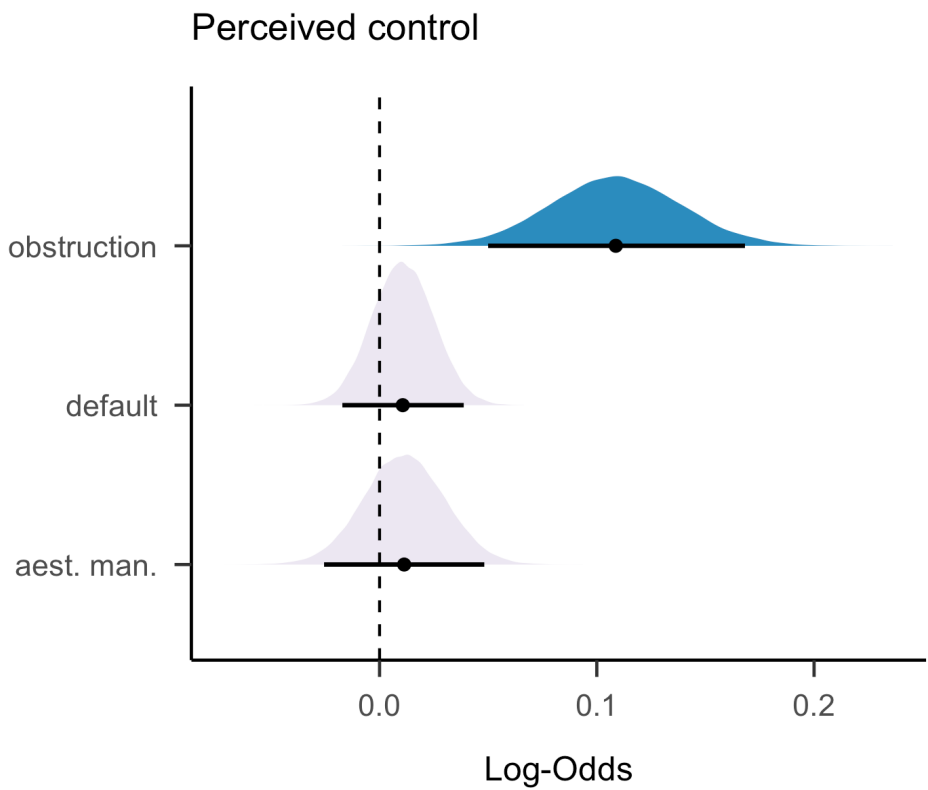

*Figure 4. Posterior distributions with mean and 95% credible interval for the predictors obstruction, default and aesthetic manipulation (outcome perceived control)*

The analysis of perceived control levels had to deal with a floor effect, meaning that a high number of observations gathered at the lower boundary of our measurement scale. This was probably partially due to how this variable was measured (i.e., slider's default position being *Not at all*), which may have led people to report generally low levels of perceived control ($M$ = 31.80, $SD$ = 28.54, Figure 5). Furthermore, we checked again whether accounting for a browser plugin installation that handles or deletes cookies changed the pattern of results, however, this was not the case.

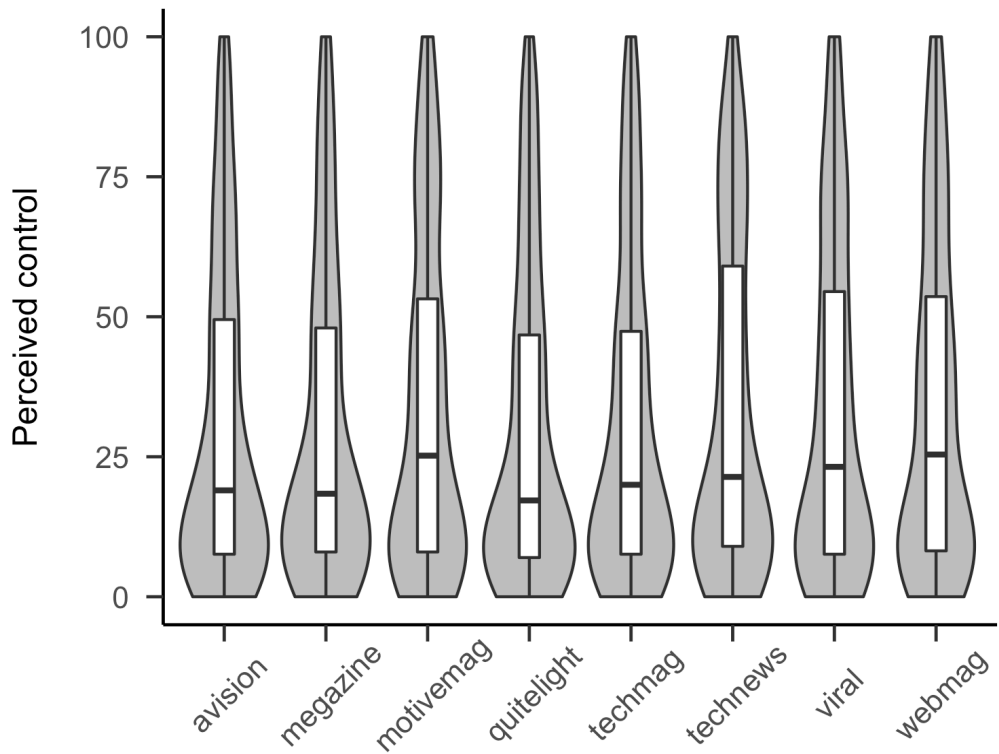

*Figure 5. Violin plots showing levels of perceived control by condition (different news websites). Grey shapes visualise the distribution of the variable, white bars represent box plots*

### 2.2.2 Exploratory Analyses

We ran two additional mixed-effects models to investigate whether the effects of the three dark patterns (default, aesthetic manipulation and obstruction) on participants' consent decisions and their perception of control depend on how much participants deliberated about their choice.

Our findings suggest that the extent to which participants deliberated about their choices did not substantially influence the effects of the three dark patterns on participants' consent decisions: default, $\beta$ = 0.19 (0.74), CrI 95% [-1.29, 1.69], OR = 1.21, aesthetic manipulation, $\beta$ = -0.31 (0.94), CrI 95% [-2.13, 1.59], OR = 0.74, and obstruction, $\beta$ = 0.79 (1.05), CrI 95% [-1.23, 3.02], OR = 2.20. Neither did the extent to which participants deliberated about their choices substantially influence the effects of the three dark patterns on participants' perceived control: default, $\beta$ = -0.02 (0.02), CrI 95% [-0.06, 0.02], OR = 0.98, aesthetic manipulation, $\beta$ = -0.01 (0.02), CrI 95% [-0.04, 0.03], OR = 0.99, and obstruction, $\beta$ = -0.01 (0.03), CrI 95% [-0.06, 0.04], OR = 0.99. This finding may be due to the fact that participants reported generally low levels of deliberation ($M$ = 20.99, $SD$ = 25.33). Similar to the perceived control measurement, absolute values of deliberation should be interpreted cautiously due to the assessment procedure (i.e., slider's default position being *Not at all*).

## 2.3 Discussion

The goal of Experiment 1 was to investigate whether dark patterns in cookie consent requests lead users to choose the privacy-unfriendly option more often than the privacy-friendly one and whether such dark patterns make people perceive less control over their personal data. Although we could show that the majority of participants always chose the privacy-unfriendly option and reported a lack of control over their personal data, we did not find clear support for those effects being due to the dark patterns. Unexpectedly, we found that obstruction led people to perceive more rather than less control over their personal data. Given the generally low levels of perceived control, which we observed across all conditions (as shown in Figure 5), more evidence is needed before making interpretations about this association.

Apart from specific effect structures, the data provided substantial ground for further insights into how people perceive consent requests and how they act on them. Most participants reported that they did not read the consent requests properly and did not think much about their decision before choosing one option. Still, the majority of participants agreed to all consent requests, seemingly in a default manner. This consent behaviour suggests that legal consent requirements for tracking cookies do not work as intended by law. At least, this conclusion applies to the way how cookie consent requests are often presented in practice. People do not seem to engage with privacy decisions in a rational and deliberate manner, as assumed by the privacy-calculus theory and, partly, by EU privacy law (GDPR, 2016, recital 7).

One reason for this observed default behaviour may be that people are conditioned to agree to consent request from their everyday life. Many websites do not even provide the opportunity to choose between different options, but make access to the site conditional on accepting tracking cookies with so-called "tracking walls" (called "forced action" by Gray et al., 2018; Zuiderveen Borgesius et al., 2017a). Hence, people often have to consent to access the content of a website or other service. It might be that the conditioned behaviour from reviewing consent requests on a daily basis overwrote the effects of the dark patterns in Experiment 1. This would be in line with the finding that people did not think much about their decision, but possibly followed the heuristic approach of choosing the option they normally choose.

To see how the design nudges relate to the observed (and possibly conditioned) default behaviour and to further investigate the unexpected effect of obstruction increasing perceived control, we conducted Experiment 2. In this follow-up experiment, we reversed the direction of the design nudges (i.e., towards the privacy-friendly option). By applying the design nudges in this "unconventional" way we aimed to see whether this would change the behaviour observed in Experiment 1. Further discussion of Experiment 1 will follow in the general discussion after Experiment 2.

## 3 EXPERIMENT 2

### 3.1 Method

As for the first experiment, we preregistered our sample size estimation, hypotheses and statistical analysis before running Experiment 2. The preregistration, the code of the study application, all used materials, data, and analysis scripts are again available on the Open Science Framework (https://osf.io/bfdvy/). Information about the used R version and all packages can be found in Appendix A. Following, we will only describe the differences between the original Experiment 1 and the follow-up Experiment 2 to avoid repetition.

#### *3.1.1 Procedure and Design*

Whereas the general procedure and design stayed the same in the follow-up experiment, we asked participants additionally about their privacy fatigue. This questionnaire was added to the second part of the study, just before we assessed general privacy concerns.

#### *3.1.2 Web application and Materials*

3.1.2.1 Consent requests

To reverse the direction of the design nudges, the focus was now on the "Do Not Agree" (to tracking) option instead of the "Agree" option. Hence, default was represented by a preselected "Do Not Agree" radio button on the websites Quitelight, Techmag, Technews and Webmag (Figure 6 shows one example consent request; screenshots of all consent requests can be found on the Open Science Framework). Aesthetic manipulation was represented by a blue coloured "Do Not Agree" button on the websites Megazine, Techmag, Viral and Webmag. Obstruction was represented by the option "Manage options" instead of "Agree" on the websites Motivemag, Technews, Viral and Webmag. Participants could only choose "Agree" after selecting "Manage options". The consent request of the website Avision represented, as in Experiment 1, the baseline condition with none of the three design nudges included.

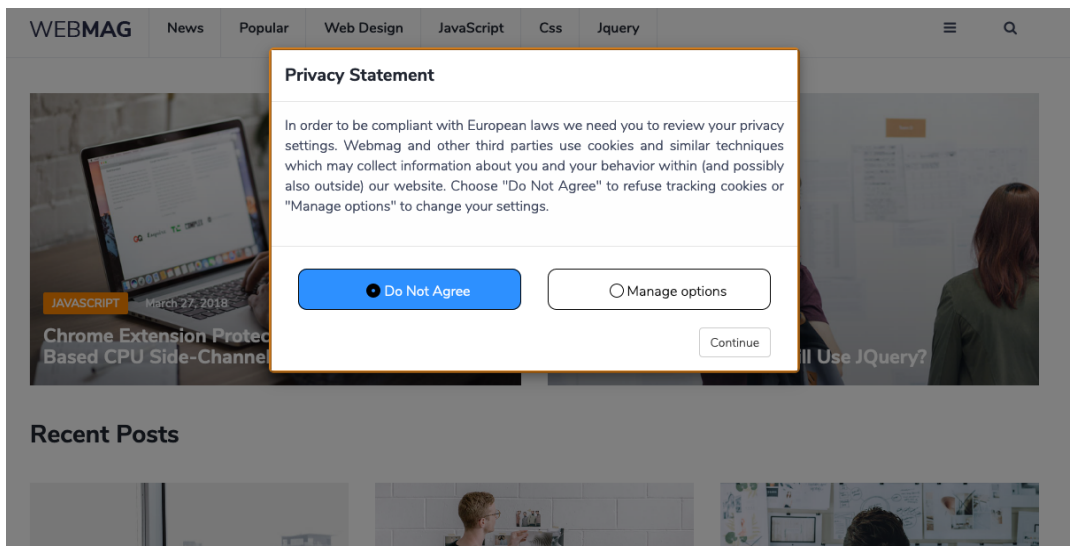

*Figure 6. Example consent request featuring all three bright patterns default, aesthetic manipulation and obstruction. Website: Webmag*

3.1.2.2 Measures

All measures of Experiment 1 were also in place in Experiment 2. As in the first study, perceived control (range: 0 - 100, $M$ = 39.55, $SD$ = 28.90) showed very good internal consistency with a raw Cronbach's $\alpha$ = 0.98. Observed deliberation levels had a range of 0 - 100, $M$ = 26.97 and $SD$ = 28.61. General privacy concerns, with a range of 1.33 - 7, $M$ = 4.23 and $SD$ = 1.15, showed acceptable to good internal consistency with a raw Cronbach's $\alpha$ = 0.77.

Manipulation checks included again the question for each consent request whether the participant had read the consent information (in 11.8% of the cases "Read it completely", 48.2% "Skimmed it", 40.0% "Did not read it at all") before clicking on one option and whether they remembered which option ("Agree", "Do Not Agree") they had chosen (15.3% of all consent decisions could not be remembered correctly). Further, participants provided information on whether they had installed a browser plugin, which handles or deletes cookies (19.8% "Yes", 80.2% "No").

Additionally to all measures of Experiment 1, we added a questionnaire about privacy fatigue, developed by Choi et al. (2018), to the follow-up experiment. We did not include items that Choi et al. (2018) deleted due to cross-loading (or other reasons). Privacy fatigue, measured on a seven-point scale ranging from *Strongly disagree* to *Strongly agree*, with a range of 1.50 - 7, $M$ = 4.40 and $SD$ = 1.01, showed acceptable internal consistency with a raw Cronbach's $\alpha$ = 0.71. Although dropping questionnaire item one would increase the overall $\alpha$-level by 0.02 we refrained from doing so because the increase was too marginal to question the theoretical structure of the scale. We used the average of all six items in the statistical analysis to form the control variable privacy fatigue.

### 3.1.3 Participants

We recruited a total of $N$ = 255 participants for Experiment 2 via the crowdsourcing platform Prolific Academic. This sample size was based on the sample size of Experiment 1 and followed the same inclusion criteria.

On average it took participants 10.60 minutes ($SD$ = 5.28) to complete the study. We left 54 participants out of this calculation because they showed very long completion times, indicating that they divided the study over several days. Yet, their consent behaviour did not seem to differ from the rest of the sample and thus they were kept for analysis. Similar as in Experiment 1, only 7 participants completed the experiment in less than 5 minutes (but not under 3 minutes). Because of the low number we kept them in the sample. We excluded participants who could not finish the study due to technical problems.

The total sample population consisted of 175 females (68.6%), 79 males (31.0%), 1 person identifying as "Other" (0.4%) and had a mean age of 35.20 years ($SD$ = 10.97). Of all 255 participants who took part in the experiment, 58 dropped out in the second part of the study (i.e., after reviewing the eight news websites). Again, none of the dropouts happened during the completion of a questionnaire (only in between) and we detected no prevalent pattern of missingness (e.g., the consent behaviour did not differ between participants with complete cases and those who would drop out later on). Hence, we found all participants' data eligible for analysis.

### 3.1.4 Data Analyses

Experiment 2 followed the same analysis approach (including the same model structures) as Experiment 1 to ensure valid one-to-one result comparison. The only addition in Experiment 2 were two extra exploratory models (following the structure of the main models) with privacy fatigue instead of privacy concerns as the control variable, to see whether this would change the pattern of results.

## 3.2 Results

### 3.2.1 Main Analyses

To investigate our third set of hypotheses 3a/b/c (stating that bright patterns will sway people towards the "Do Not Agree" option) we again first visualise the recorded consent decisions for each news website (see Figure 7). We observed that in Experiment 2 only in slightly more than half of the cases (53.2%) people chose to agree to the consent requests, representing a reduction of 40.7% compared to Experiment 1. This time, more than one-third of the participants (36.1%) changed their consent behaviour between conditions. This trend is reflected by our results, which showed that two of the three tested design nudges swayed participants effectively towards the "Do Not Agree" option.

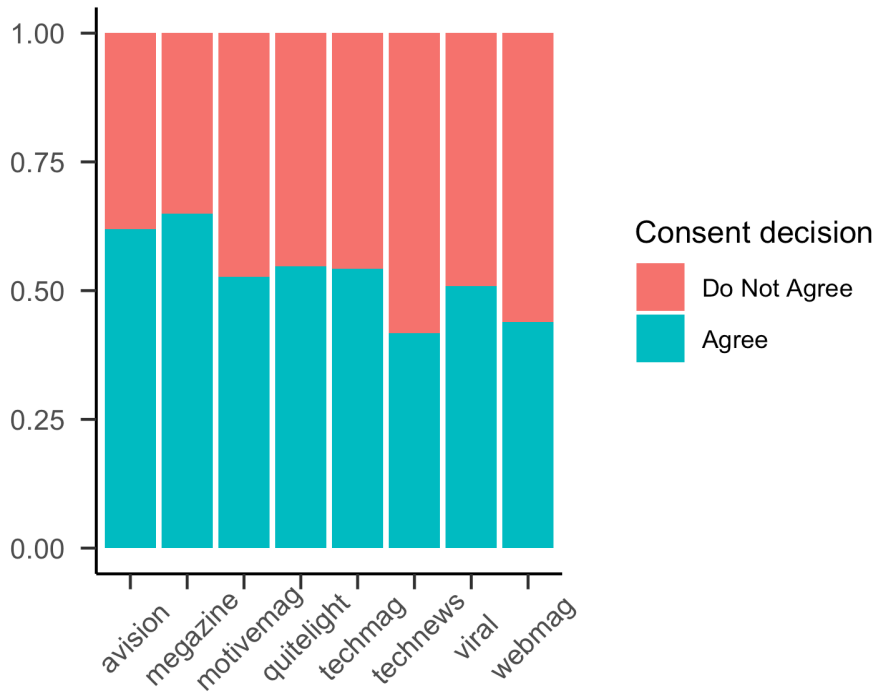

*Figure 7. Consent decisions (proportional) by condition (different news websites)*

Specifically, we found a substantial main effect of default, $\beta$ = -0.75 (0.13), CrI 95% [-1.01, -0.51], OR = 0.47, and obstruction, $\beta$ = -0.97 (0.20), CrI 95% [-1.39, -0.60], OR = 0.38, on the outcome consent decision (see Figure 8), supporting our hypotheses H3a and H3c respectively. Given that we kept the outcome consent decision coded as in the original study (0 = "Do Not Agree", 1 = "Agree"), a negative effect estimate means an increased likelihood of selecting "Do Not Agree". To interpret odds ratios which are smaller than 1 in a meaningful way we will inverse them (1/OR). Hence, if the option "Do Not Agree" was selected by default, the odds of participants choosing this option were two times higher than if the option had not been preselected. Similarly, if the "Agree" option was obstructed the odds of participants choosing the "Do Not Agree" option were two and a half times higher than if the "Agree" option had not been obstructed. We did not find support for Hypothesis H3b however, as there was no notable effect of aesthetic manipulation, $\beta$ = 0.06 (0.14), CrI 95% [-0.21, 0.34], OR = 1.06. The pattern of results did not change when additionally accounting for a participant's previous consent decision or whether the participant had a browser plugin installed that handles or deletes cookies.

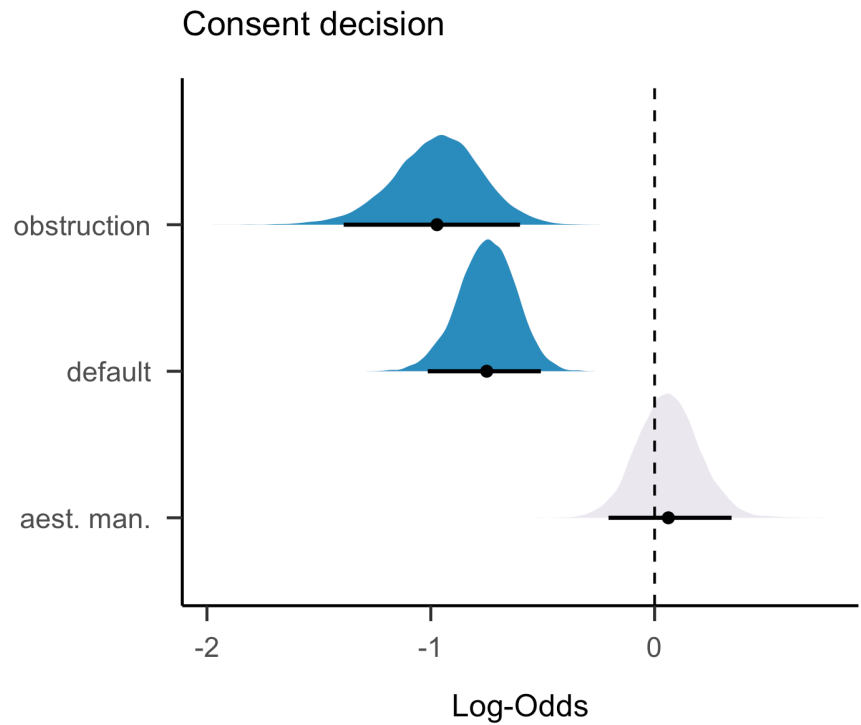

*Figure 8. Posterior distributions with mean and 95% credible interval for the predictors default, aesthetic manipulation and obstruction (outcome consent decision)*

Testing our fourth set of hypotheses 4a/b/c, we did not find that (bright) design nudges made people perceive less control over their personal data. Rather, we replicated the result pattern of Experiment 1, including the finding that obstructing one choice option led participants to report more rather than less perceived control.

Specifically, obstruction showed a small but notable main effect, $\beta$ = 0.06 (0.03), CrI 95% [0.00, 0.12], OR = 1.06, on the outcome perceived control (see Figure 9). Hence, hypothesis H4c was not supported. Further, we did not find support for hypotheses H4a and H4b concerning the effects of default, $\beta$ = -0.01 (0.01), CrI 95% [-0.04, 0.02], OR = 0.99, and aesthetic manipulation, $\beta$ = -0.02 (0.02), CrI 95% [-0.07, 0.02], OR = 0.98. We checked again whether accounting for a browser plugin installation that handles or deletes cookies changed the pattern of results, but this was not the case.

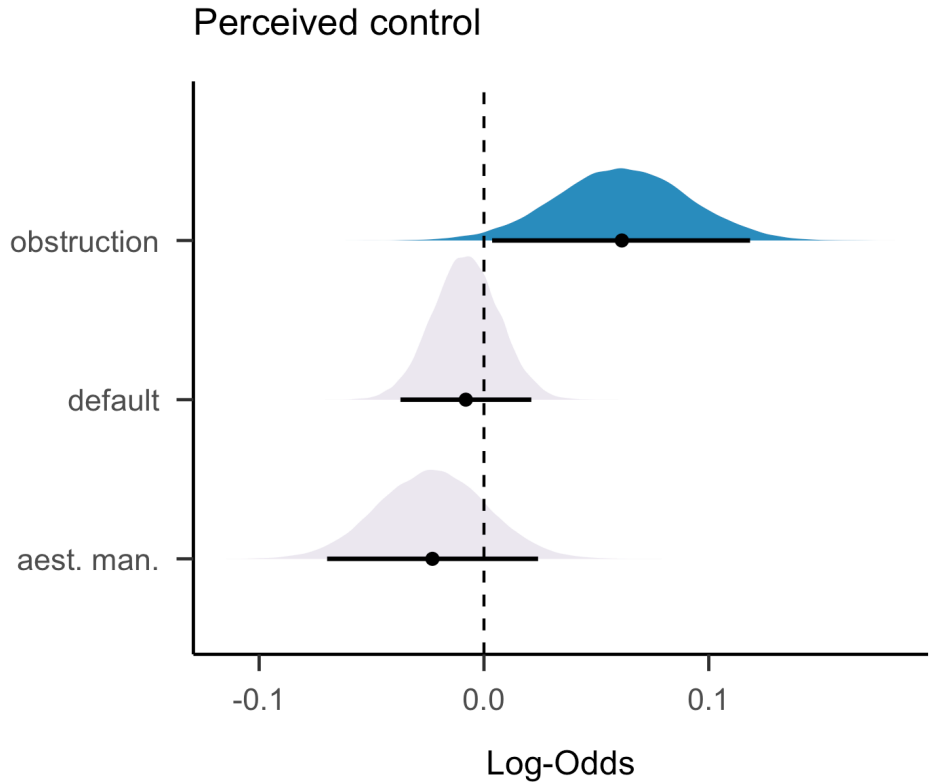

*Figure 9. Posterior distributions with mean and 95% credible interval for the predictors default, aesthetic manipulation and obstruction (outcome perceived control)*

*3.2.2    Exploratory Analyses*

Apart from investigating possible moderation effects of deliberation (as in Experiment 1), we ran two additional mixed-effects models with privacy fatigue instead of privacy concerns as a control variable to see whether this would change our results.

As in the original study, the extent to which participants deliberated about their choices did not substantially influence the effects of the three bright patterns on participants' consent decisions: default, $\beta$ = 0.05 (0.16), CrI 95% [-0.27, 0.36], OR = 1.05, aesthetic manipulation, $\beta$ = 0.02 (0.17), CrI 95% [-0.32, 0.36], OR = 1.02, and obstruction, $\beta$ = 0.22 (0.21), CrI 95% [-0.20, 0.65], OR = 1.25. Neither did the extent to which participants deliberated about their choices substantially influence the effects of the three bright patterns on participants' perceived control: default, $\beta$ = 0.00 (0.02), CrI 95% [-0.03, 0.04], OR = 1.00, aesthetic manipulation, $\beta$ = 0.01 (0.02), CrI 95% [-0.03, 0.06], OR = 1.01, and obstruction, $\beta$ = -0.01 (0.03), CrI 95% [-0.06, 0.04], OR = 0.99. This finding may be due to the fact that participants reported again generally low levels of deliberation ($M$ = 20.99, $SD$ = 25.33).

Replacing privacy concerns with privacy fatigue as the control variable showed that privacy fatigue acted across all models in the opposite direction as privacy concerns. While higher levels of privacy concerns were associated with being less likely to agree to the consent requests, higher levels of privacy fatigue were associated with being more likely to choose "Agree". However, the pattern of results did not differ whether privacy concerns or privacy fatigue was used as a control variable.

## 3.3 Discussion

The follow-up experiment aimed to explore how design nudges towards the privacy-friendly option (i.e., bright patterns) would influence people's consent choices and their perception of control compared to what we found in Experiment 1.

The first finding was that people did not agree to every cookie consent statement in a default manner anymore (as many had done in Experiment 1). Compared to Experiment 1, about ten times more people changed their consent behaviour between conditions in Experiment 2. Given that all we changed between Experiment 1 and Experiment 2 was the direction of the design nudges, the results illustrate that these seemingly small tweaks in the interface can heavily influence people's privacy choices. The results also support the suspicion that the effects of the dark patterns in Experiment 1 were obscured, because people may be conditioned to always agree to cookie consent requests. People might have developed such automatic behaviour by reviewing thousands of ambiguous cookie consent requests, or even take-it-or-leave-it choices. Specifically, we found that people were substantially more likely to choose "Do Not Agree" if this option was preselected or the alternative "Agree" option obstructed.

The result pattern regarding perceived control over one's personal data was very similar between the two experiments. In each experiment, people reported that they perceived little control over their personal data. However, we could not find that the design nudges had led to this low level of perceived control. Surprisingly, in both cases, obstructing one choice option led people to perceive more rather than less control over their personal data. One possible explanation could be that the phrase "Manage options", which obstructed either the "Agree" (Experiment 1) or "Dot Not Agree" (Experiment 2) option, conveyed somehow the feeling of control. Further discussion of these findings follows in the subsequent general discussion.

## 4      GENERAL DISCUSSION

In the last part of this paper, we first summarise our findings and discuss how they fit into the existing literature and the theoretical framework of privacy decision making (in the context of design nudges). In a second step, we shift towards practical approaches to address the problems of the current consent system. Specifically, we explore how policymakers could address problems with the legal consent requirement for tracking cookies. Lastly, we will address limitations of our experiments and give suggestions for future work.

## 4.1  Summary and theoretical implications

Overall, our findings were in line with previous research (Machuletz & Böhme, 2019; Utz et al., 2019) and form additional evidence supporting the persuasive power of design nudges on users consent choices. In Experiment 1 (featuring dark patterns that people are used to) it did not seem to differ for participants' consent behaviour whether design nudges were used or not. However, in Experiment 2 (featuring bright patterns), two out of the three tested design nudges substantially affected people's consent choices in the hypothesised direction. As the only difference between the two experiments was the direction of the design nudges, it appears that such nudges influence privacy choices after all.

Why did we observe this discrepancy between the results of the two experiments? Nudges are often thought of as manipulations of the choice environment which only elicit their potential effect while being in place (i.e., no long-term effect). However, it may be that this changes when nudges (specifically System 1 nudges) are used for longer periods of time (e.g., seeing consent requests with dark patterns for years). A form of conditioning may happen, ultimately leading people to behave in a certain way even in absence of the nudge (e.g., participants agreeing to the consent request in the baseline condition without any design nudges present). Hertwig and Grüne-Yanoff (2017) refer to this process of "effect survival" after the removal of the nudge as the development of behavioural routines. Of course, design nudges are probably not the only reason for this conditioning to happen, but they certainly have the potential to play an important role.

Concerning the influence of the design nudges on participants' perception of control over their personal data, our results were stable across both experiments but did not support our assumptions. Although participants had (theoretically) full control over each decision in our study (i.e., for each consent request there was the possibility to choose "Do Not Agree"), they did not seem to perceive it that way, possibly because they are used to ambiguous real-life consent requests, which do not always offer a meaningful choice. Surprisingly, people perceived more rather than less control if one choice option was hidden behind "Manage options". As mentioned in the discussion of Experiment 2 the formulation "Manage options" may have somehow (unjustified) conveyed the feeling of control, highlighting the manipulative effect of design nudges. This is in line with what Forbrukerrådet (2018) describe as the "illusion of control". Further considerations and suggestions for future work are discussed at the end of the paper.

## 4.2 Practical implications

Taking these findings into account, the question arises how problems of the current consent system for tracking cookies could be addressed so that online privacy self-management works in a meaningful way. We believe that there are two ways to tackle the issue: Focusing on users or on companies. First, we discuss approaches that focus on the user.

By focusing on the user, we mean any attempt to change the behaviour or competences of the user. Following Hertwig and Grüne-Yanoff (2017), we differentiate between nudging approaches, which try to change behaviour by altering the choice architecture, and boosting approaches, which focus on competence building to enable a certain behaviour. Non-educative nudges, such as bright patterns, could be used to nudge users towards the privacy-friendly option, as in Experiment 2. These bright patterns do not require any motivation from the user but may lead to similar problems as their dark counterparts, such as unreflective default behaviour and users' perception of a lack of control.

Further, there are educative nudges (after Sunstein, 2016b) such as reminders or warnings, which build a middle ground between nudging and boosting, because they require some level of motivation to foster a context-specific competence (called short-term boosts by Hertwig & Grüne-Yanoff, 2017). In the context of cookie consent requests, an example of an educative nudge could be feedback about possible consequences of a choice. However, the company that asks for consent would have to implement this educative nudge. As many companies have incentives to nudge internet users towards the privacy-unfriendly option (e.g., to collect data for targeted advertising), the practical feasibility of such nudges is questionable. After all, if policymakers require companies to implement pro privacy nudges, the companies can sabotage those nudges (Willis, 2014).

Lastly, there are long-term boosts, which aim at a permanent change of skills and decision tools. In theory, boosts, which aim at building procedural rules such as "When I see a consent request I read the provided information before making a choice" could be used in the context of cookie consent requests. Such boosts, in theory, could be suitable to break people out of automatic behaviour and to help them deliberate before making a choice. However, long-term boosts are often costlier than nudges (e.g., changing a default requires less time and effort than creating an intervention to form procedural rules). In addition, boosts only work if people are motivated to acquire new skills.

Presumably, people's motivation to deliberate about cookie consent requests is low. If somebody wants to visit a website, having to think about a consent request is an unwelcome hurdle. If people lack the motivation to build certain competences, Hertwig (2017) advises to use nudging rather than boosting approaches. This brings us back to bright patterns, which do not require motivation from the user. However, as noted, many companies using dark patterns have an interest in tracking people's online behaviour, so it does not seem plausible that such companies will implement effective pro-privacy nudges. Consequently, user-focused approaches seem unrealistic for the context of cookie consent requests.

A second strategy focuses not on the user, but on changing the behaviour of companies. Amending legal requirements can influence company behaviour. Our consent requests were designed in a way that they resemble many of those requests used in practice under the ePrivacy Directive. Thus, the results of our experiments illustrate that consent requests often do not lead to genuinely "informed" consent, considering that most participants did not read the consent information, and reported a lack of control over their personal data. Dark patterns may play a role in that, but based on our study findings it cannot be concluded that stricter design regulations for consent requests alone (i.e., banning dark patterns from consent requests) would resolve the problem. After all, most participants also agreed to web tracking in the baseline condition of Experiment 1 without any design nudge present. Overall, this study contributes to a body of research that questions the effectiveness of legal informed consent requirements as a privacy protection tool (Acquisti et al., 2017; Zuiderveen Borgesius, 2015a). How should policymakers react?

Could enforcement of current law push companies to use bright patterns? As noted, the ePrivacy Directive (2009) requires consent for tracking cookies and similar tracking techniques; the GDPR's strict conditions for valid consent apply. But these two instruments do not explicitly ban dark patterns in consent requests, let alone require bright patterns. Dark patterns do violate the spirit of the GDPR, for two reasons. First, the GDPR requires that personal data are only collected "fairly and in a transparent manner in relation to the data subject" (article 5(1)(a)). Many dark patterns could be regarded as unfair. However, the fairness requirement is rather vague, and therefore difficult to enforce.

Second, an argument could be made that the GDPR generally discourages the use of dark patterns, because the GDPR bans certain types of dark patterns. For instance, the GDPR bans opt-out systems (that assume consent if people fail to object), pre-selected "I consent" options, and certain types of tracking walls and similar take-it-or-leave-it choices (article 4(11) and article 7). The GDPR also states that a consent "request must be clear, concise and not unnecessarily disruptive" (recital 32), and must use "plain language and (…) should not contain unfair terms" (recital 42). Some dark patterns may violate those requirements. Moreover, European regulators note that "dark patterns (…) are contrary to the spirit of Article 25" of the GDPR, which requires privacy by design (European Data Protection Board, 2020).

All in all, the extent to which the GDPR bans dark patterns must become clear in case law and enforcement actions by Data Protection Authorities. In 2018, seven consumer organisation filed complaints with national Data Protection Authorities regarding location tracking by Google. The organisations also complain about dark patterns (BEUC, 2020). However, Data Protection Authorities did not finish their investigations yet. More generally, it may take a long time before there is enough case law to push companies towards abandoning dark patterns.

Amendments to the law could be useful. The European Commission (2017) published a proposal to replace the ePrivacy Directive with an ePrivacy Regulation. The proposal for the ePrivacy Regulation contains promising ideas, especially after the European Parliament (2017) amended it. For instance, under the ePrivacy Regulation, it would be obligatory for any company to respect "Do Not Track" and similar signals (European Parliament, 2017). With "Do Not Track" or a similar system, an internet user can choose a setting on their device once, which communicates to all websites and tracking companies that the user does not want to be tracked. Such a "Do Not Track"-like solution could limit the number of times that people are asked to consent to tracking (Zuiderveen Borgesius et al., 2017).

Perhaps additional rules are needed to ensure that companies refrain from asking people to make an exception to their "do not track me" setting. The ePrivacy proposal also bans companies from using "tracking walls", a barrier that visitors can only pass if they consent to tracking by third parties (European Parliament, 2017). However, at the moment it is unclear whether and in what form the ePrivacy proposal will be adopted (Legislative Train Schedule, 2020).

### 4.3 Limitations and suggestions for future research

The first limitation of our study that future research should address is the location of the presented choice options. Known as Fitts's law, which is a predictive model of human movement, one can assume that it is easier and faster to hit larger targets closer to you than smaller targets further away from you (MacKenzie, 1992). In our design of the consent requests the "Agree" option was on the right-hand side and thus closer to the "Continue" button (which was also on the right-hand side) than the "Do Not Agree" option, which was on the left-hand side. This setup was inspired by what we saw in real-life practice but might have acted as an additional design nudge. It could be interesting to include eye-tracking measurements to follow participants visual attention while they encounter cookie consent requests.

A second limitation relates to our conceptualisation of rational choice. We base ourselves on the privacy calculus theory to weigh the privacy risks against the privacy benefits of each choice in the consent request. Not included in this calculation are factors such as little time differences between choosing one option versus the other, which arise for instance when one option is obstructed (e.g., choosing one option requires more mouse clicks than the other). These factors, however, are often the mechanistic core of a design nudge and thus hard to "strip away".

Third, we had to compromise between ecological validity and a controlled experimental setting for the design of our consent requests. To include all three design nudges at the same time, we had to choose a consent request setup, which deviated slightly from most real-life consent requests. Namely, we presented the available choice possibilities in the form of radio buttons (which can be ticked) instead of clickable buttons, because regular buttons cannot be preselected (which is needed for the design nudge default).

Fourth, we had to tweak some aspects of the design of each consent request (see Appendix B or the Open Science Framework) to match the design of the corresponding news website and make the cover story of eight independent external news websites plausible. While these changes may seem arbitrary, we paid close attention to not change any parts close to the choice options in which our manipulations where applied.

Fifth, our design complicated the reliable measurement of participants' perceived control over their personal data, which was assessed with a time delay to the actual consent decisions (i.e., after all eight news websites had been reviewed). This was due to our study design involving the cover story about the first impression of the design of news websites, which would have been compromised when drawing attention on the consent requests during part 1 of the experiment (i.e., while reviewing the news websites). In addition, the slider with which people could indicate how much control they perceived had a default setting of *Not at all*. We chose this setting because it resembled in our opinion the most neutral and intuitively understandable starting position (compared to the middle between *Not at all* and *Complete control*). However, this setting may have partially caused the previously discussed floor effect (see results section of Experiment 1). Future studies should reconsider the scale's default position and its possible consequences for measurement. Nonetheless, we hope to have created a starting point for future research to assess perceived control specifically in the context of consent requests. Further, it may be valuable to investigate the concept of perceived control additionally through a qualitative approach to shed light onto the possible shortcomings of the quantitative approach which was used so far.

Lastly, future research should investigate whether and under what circumstances conditioning and behavioural routines develop regarding informed consent procedures. In a second step, it could be examined how these behavioural routines may be disrupted, for instance by applying friction to the decision process to stimulate deliberation (Terpstra, Schouten, Rooij, & Leenes, 2019; Zuiderveen Borgesius, 2015).

### 4.4 Conclusion

Overall, this project shed light on some of the mechanisms of design nudges in cookie consent requests. Our research findings demonstrate some of the shortcomings of legal consent requirements for cookies and similar rules that expect people to make many informed choices about their privacy. We explored possible solutions to face these shortcomings. For instance, the upcoming ePrivacy Regulation of the EU should limit the number of cookie consent requests people are confronted with. Policymakers should not put unreasonable burdens on people's shoulders and avoid responsibilisation. Responsibilisation describes "the process whereby subjects are rendered individually responsible for a task which previously would have been the duty of another – usually a state agency – or would not have been recognized as a responsibility at all" (Wakefield & Fleming, 2009, p. 276; see also Gürses, 2014). In conclusion, the concept of informed consent is not obsolete in the digital era but should be used wisely and sparingly (see also Böhme and Köpsell, 2010). In the case of web tracking and personalisation, this could mean, for instance, a global option in the browser which has to be set only once.

## FUNDING STATEMENT

This project is primarily funded by the Behavioural Science Institut (BSI) and partly by the Dutch Research Council (NWO), for the research program 'SocialMovez' with project number VI.C.181.045

## ACKNOWLEDGEMENTS

We would like to thank the reviewers for their valuable comments to improve the manuscript. Further, we thank Franc Grootjen and Ayke van Laethem for their technical advice; Franc especially for providing a university server.

## REFERENCES

Acquisti, A., Sleeper, M., Wang, Y., Wilson, S., Adjerid, I., Balebako, R., … Schaub, F. (2017). Nudges for privacy and security. ACM Computing Surveys, 50(3), 1–41. https://doi.org/10.1145/3054926

Albar, F. M., & Jetter, A. J. (2009). Heuristics in decision making. In PICMET '09 - 2009 Portland International Conference on Management of Engineering & Technology (pp. 578–584). IEEE. https://doi.org/10.1109/PICMET.2009.5262123

An, N. Z. (2019). Multi-step modals for Bootstrap. Retrieved from https://github.com/ngzhian/multi-step-modal

Archer, M. S. (2013). Rational choice theory. Routledge. https://doi.org/10.4324/9780203133897

Auguie, B. (2017). GridExtra: Miscellaneous functions for "grid" graphics. Retrieved from https://CRAN.R-project.org/package=gridExtra

Aust, F., & Barth, M. (2020). papaja: Create APA manuscripts with R Markdown. Retrieved from https://github.com/crsh/papaja

Awad, N. F., & Krishnan, M. S. (2006). The personalization privacy paradox: An empirical evaluation of information transparency and the willingness to be profiled online for personalization. MIS Quarterly, 1328.

Barr, D. J., Levy, R., Scheepers, C., & Tily, H. J. (2013). Random effects structure for confirmatory hypothesis testing: Keep it maximal. Journal of

Memory and Language, 68(3), 255–278. https://doi.org/10.1016/j.jml.2012.11.001

BEUC. (2020). The long and winding road. Two years of the GDPR: A cross-border data protection enforcement case from a consumer perspective. Retrieved from https://www.beuc.eu/publications/beuc-x-2020-074_two_years_of_the_gdpr_a_cross-border_data_protection_enforcement_case_from_a_consumer_perspective.pdf

Böhme, R., & Köpsell, S. (2010). Trained to accept?: A field experiment on consent dialogs. In Proceedings of the 28th international conference on Human factors in computing systems - CHI '10 (p. 2403). Atlanta, Georgia, USA: ACM Press. https://doi.org/10.1145/1753326.1753689

Bösch, C., Erb, B., Kargl, F., Kopp, H., & Pfattheicher, S. (2016). Tales from the dark side: Privacy dark strategies and privacy dark patterns. Proceedings on Privacy Enhancing Technologies, 2016(4), 237–254. https://doi.org/10.1515/popets-2016-0038

Brignull, H. (n.d.). Dark patterns. Retrieved from https://darkpatterns.org/

Brooke, B. (2011). Browser back button detection. Retrieved from http://www.bajb.net/2010/02/browser-back-button-detection/

Browne, W. J., & Draper, D. (2006). A comparison of Bayesian and likelihood-based methods for fitting multilevel models. Bayesian Analysis, 1(3), 473–514. https://doi.org/10.1214/06-BA117

Bryan, M. L., & Jenkins, S. P. (2016). Multilevel modelling of country effects: A cautionary tale. European Sociological Review, 32(1), 3–22. https://doi.org/10.1093/esr/jcv059

Bürkner, P.-C. (2017). brms: An R package for Bayesian multilevel models using Stan. Journal of Statistical Software, 80(1), 1–28. https://doi.org/10.18637/jss.v080.i01

Bürkner, P.-C. (2018). Advanced Bayesian multilevel modeling with the R package brms. The R Journal, 10(1), 395–411. https://doi.org/10.32614/RJ-2018-017

Carpenter, B., Gelman, A., Hoffman, M., Lee, D., Goodrich, B., Betancourt, M., … Riddell, A. (2017). Stan: A probabilistic programming language. Journal of Statistical Software, Articles, 76(1), 1–32. https://doi.org/10.18637/jss.v076.i01

Choi, H., Park, J., & Jung, Y. (2018). The role of privacy fatigue in online privacy behavior. Computers in Human Behavior, 81, 42–51. https://doi.org/10.1016/j.chb.2017.12.001

Colorbib. (2019). 28 best free news website templates 2019. Colorlib. Retrieved from https://colorlib.com/wp/free-news-website-templates/

Dijksterhuis, A., Bos, M. W., Nordgren, L. F., & van Baaren, R. B. (2006). On making the right choice: The deliberation-without-attention effect. Science, 311(5763), 1005–1007. https://doi.org/10.1126/science.1121629

Eddelbuettel, D., & Balamuta, J. J. (2017). Extending extitR with extitC++: A Brief Introduction to extitRcpp. PeerJ Preprints, 5, e3188v1. https://doi.org/10.7287/peerj.preprints.3188v1

Eddelbuettel, D., & François, R. (2011). Rcpp: Seamless R and C++ integration. Journal of Statistical Software, 40(8), 1–18. https://doi.org/10.18637/jss.v040.i08

ePrivacy Directive. (2009). Directive 2002/58/EC of the European Parliament and of the Council of 12 July 2002 concerning the processing of personal data and the protection of privacy in the electronic communications sector (directive on privacy and electronic communications), last amended by Directive 2009/136/EC of the European Parliament and of the Council of 25 November 2009 (OJ L 337 11). Retrieved from https://eur-lex.europa.eu/eli/dir/2002/58/2009-12-19

European Commission. (2017). Proposal for a regulation of the European Parliament and of the Council, concerning the respect for private life and the protection of personal data in electronic communications and repealing Directive 2002/58/EC (Regulation on Privacy and Electronic Communications) (No. COM/2017/010 final - 2017/03 (COD)). Retrieved from https://eur-lex.europa.eu/legal-content/EN/TXT/?uri=celex:52017PC0010

European Data Protection Board. (2020). Guidelines 4/2019 on Article 25 data protection by design and by default version 2.0, adopted on 20 October 2020. Retrieved from https://edpb.europa.eu/sites/edpb/files/files/file1/edpb_guidelines_201904_dataprotection_by_design_and_by_default_v2.0_en.pdf

European Parliament. (2017). Draft European Parliament Legislative Resolution on the proposal for a regulation of the European Parliament and of the Council concerning the respect for private life and the protection of personal data in electronic communications and repealing Directive 2002/58/EC (Regulation on Privacy and Electronic Communications) (No. COM(2017)0010 C8-0009/2017 2017/0003(COD)). Retrieved from https://www.europarl.europa.eu/doceo/document/A-8-2017-0324_EN.html

Fansher, M., Chivukula, S. S., & Gray, C. M. (2018). #Darkpatterns. In R. Mandryk, M. Hancock, M. Perry, & A. Cox (Eds.), Extended Abstracts of the 2018 CHI Conference on Human Factors in Computing Systems - CHI '18 (pp. 1–6). New York, New York, USA: ACM Press. https://doi.org/10.1145/3170427.3188553

Ferrari, S., & Cribari-Neto, F. (2004). Beta regression for modelling rates and proportions. Journal of Applied Statistics, 31(7), 799–815. https://doi.org/10.1080/0266476042000214501

Forbrukerrådet. (2018). Deceived by design: How tech companies use dark patterns to discourage us from exercising our rights to privacy. Retrieved from https://www.forbrukerradet.no/undersokelse/no-undersokelsekategori/deceived-by-design/

GDPR. (2016). Regulation (EU) 2016/679 of the European Parliament and of the Council of 27 April 2016 on the protection of natural persons with regard to the processing of personal data and on the free movement of such data, and repealing Directive 95/46/EC (General Data Protection Regulation). Official Journal L, 119, 1–88. Retrieved from https://eur-lex.europa.eu/eli/reg/2016/679/oj

Gray, C. M., Kou, Y., Battles, B., Hoggatt, J., & Toombs, A. L. (2018). The dark (patterns) side of UX design. In R. Mandryk, M. Hancock, M. Perry, & A. Cox (Eds.), Proceedings of the 2018 CHI Conference on Human Factors in Computing Systems - CHI '18 (pp. 1–14). New York, New York, USA: ACM Press. https://doi.org/10.1145/3173574.3174108

Grosjean, P., & Ibanez, F. (2018). Pastecs: Package for analysis of space-time ecological series. Retrieved from https://CRAN.R-project.org/package=pastecs

Gürses, S. (2014). Attitudes towards "Spiny CACTOS". Retrieved from https://vous-etes-ici.net/next-week-spiny-cactos-at-usec-2014/

Hertwig, R. (2017). When to consider boosting: Some rules for policy-makers. Behavioural Public Policy, 1(02), 143–161. https://doi.org/10.1017/bpp.2016.14

Hertwig, R., & Grüne-Yanoff, T. (2017). Nudging and boosting: Steering or empowering good decisions. Perspectives on Psychological Science : A Journal of the Association for Psychological Science, 12(6), 973–986. https://doi.org/10.1177/1745691617702496

Kahneman, D. (2011). Thinking, fast and slow (1st ed). New York: Farrar, Straus and Giroux.

Kay, M. (2020). tidybayes: Tidy data and geoms for Bayesian models. https://doi.org/10.5281/zenodo.1308151

Kowarik, A., & Templ, M. (2016). Imputation with the R package VIM. Journal of Statistical Software, 74(7), 1–16. https://doi.org/10.18637/jss.v074.i07

Lai, Y.-L., & Hui, K.-L. (2006). Internet opt-in and opt-out: Investigating the roles of frames, defaults and privacy concerns. In Proceedings of the 2006 ACM SIGMIS CPR conference on computer personnel research Forty four years of computer personnel research: Achievements, challenges & the future - SIGMIS CPR '06 (p. 253). Claremont, California, USA: ACM Press. https://doi.org/10.1145/1125170.1125230

Laufer, R. S., & Wolfe, M. (1977). Privacy as a concept and a social issue: A multidimensional developmental theory. Journal of Social Issues, 33(3), 22–42. https://doi.org/10.1111/j.1540-4560.1977.tb01880.x

Legislative Train Schedule. (2020). Proposal for a regulation on privacy and electronic communications. Retrieved from

https://www.europarl.europa.eu/legislative-train/theme-connected-digital-single-market/file-jd-e-privacy-reform

Lord, D., Mönnich, A., Ronacher, A., & Unterwaditzer, M. (2010). Flask (a Python microframework). Retrieved from http://flask.pocoo.org/

Luguri, J., & Strahilevitz, L. (2019). Shining a light on dark patterns. SSRN Electronic Journal. https://doi.org/10.2139/ssrn.3431205

Machuletz, D., & Böhme, R. (2019). Multiple purposes, multiple problems: A user study of consent dialogs after GDPR. arXiv:1908.10048 [Cs]. Retrieved from http://arxiv.org/abs/1908.10048

MacKenzie, I. S. (1992). Fitts' Law as a research and design tool in Human-Computer Interaction. HumanComputer Interaction, 7(1), 91–139. https://doi.org/10.1207/s15327051hci0701_3

Malhotra, N. K., Kim, S. S., & Agarwal, J. (2004). Internet users' information privacy concerns (IUIPC): The construct, the scale, and a causal model. Information Systems Research, 15(4), 336–355. https://doi.org/10.1287/isre.1040.0032

Morey, R. D., Hoekstra, R., Rouder, J. N., Lee, M. D., & Wagenmakers, E.-J. (2016). The fallacy of placing confidence in confidence intervals. Psychonomic Bulletin & Review, 23(1), 103–123. https://doi.org/10.3758/s13423-015-0947-8

Mullen, L. A., Benoit, K., Keyes, O., Selivanov, D., & Arnold, J. (2018). Fast, consistent tokenization of natural language text. Journal of Open Source Software, 3(23), 655. https://doi.org/10.21105/joss.00655

Müller, K. (2017). Here: A simpler way to find your files. Retrieved from https://CRAN.R-project.org/package=here

Nouwens, M., Liccardi, I., Veale, M., Karger, D., & Kagal, L. (2020). Dark patterns after the GDPR: Scraping consent pop-ups and demonstrating their influence. arXiv:2001.02479 [Cs]. https://doi.org/10.1145/3313831.3376321

R Core Team. (2020). R: A language and environment for statistical computing. Vienna, Austria: R Foundation for Statistical Computing. Retrieved from https://www.R-project.org/

Revelle, W. (2019). Psych: Procedures for psychological, psychometric, and personality research. Evanston, Illinois: Northwestern University. Retrieved from https://CRAN.R-project.org/package=psych

Schubert, C. (2015). On the ethics of public nudging: Autonomy and agency. SSRN Electronic Journal. https://doi.org/10.2139/ssrn.2672970

Simon, H. A. (1957). Models of man, social and rational: Mathematical essays on rational human behavior in a social setting. New York, NY, USA: Wiley.

Smith, H. J., Dinev, T., & Xu, H. (2011). Information privacy research: An interdisciplinary review. MIS Quarterly, 35(4), 989–1015.

Stauffer, R., Mayr, G. J., Dabernig, M., & Zeileis, A. (2009). Somewhere over the rainbow: How to make effective use of colors in meteorological visualizations. Bulletin of the American Meteorological Society, 96(2), 203–216. https://doi.org/10.1175/BAMS-D-13-00155.1

Sunstein, C. R. (2016a). People prefer system 2 nudges (kind of). SSRN Electronic Journal. https://doi.org/10.2139/ssrn.2731868

Sunstein, C. R. (2016b). The ethics of influence: Government in the age of behavioral science. Cambridge University Press.

Terpstra, A., Schouten, A. P., Rooij, A. de, & Leenes, R. E. (2019). Improving privacy choice through design: How designing for reflection could support privacy self-management. First Monday, 24(7). https://doi.org/10.5210/fm.v24i7.9358

Thaler, R. H. (2018). Nudge, not sludge. Science, 361(6401), 431–431. https://doi.org/10.1126/science.aau9241

Thaler, R. H., & Sunstein, C. R. (2009). Nudge: Improving decisions about health, wealth, and happiness (Rev. and expanded ed). New York: Penguin Books.

Utz, C., Degeling, M., Fahl, S., Schaub, F., & Holz, T. (2019). (Un)Informed consent: Studying GDPR consent notices in the field. In ACM SIGSAC Conference on Computer and CommunicationsSecurity (CCS '19) (p. 18). London, United Kingdom. Retrieved from https://arxiv.org/pdf/1909.02638.pdf

Wakefield, A., & Fleming, J. (2009). The Sage dictionary of policing. Los Angeles; London: SAGE. Retrieved from http://www.dawsonera.com/depp/reader/protected/external/AbstractView/S9781446207017

Wickham, H. (2011). The split-apply-combine strategy for data analysis. Journal of Statistical Software, 40(1), 1–29. Retrieved from http://www.jstatsoft.org/v40/i01/

Wickham, H. (2016). Ggplot2: Elegant graphics for data analysis. Springer-Verlag New York. Retrieved from https://ggplot2.tidyverse.org

Wickham, H. (2019). Stringr: Simple, consistent wrappers for common string operations. Retrieved from https://CRAN.R-project.org/package=stringr

Wickham, H., François, R., Henry, L., & Müller, K. (2020). Dplyr: A grammar of data manipulation. Retrieved from https://CRAN.R-project.org/package=dplyr

Wickham, H., & Henry, L. (2020). Tidyr: Tidy messy data. Retrieved from https://CRAN.R-project.org/package=tidyr

Willis, L. E. (2014). Why not privacy by default. Berkeley Technology Law Journal, 29, 61. Retrieved from https://heinonline.org/HOL/Page?handle=hein.journals/berktech29\&id=71\&div=\&collection=

Xie, Y. (2015). Dynamic documents with R and knitr (2nd ed.). Boca Raton, Florida: Chapman; Hall/CRC. Retrieved from https://yihui.org/knitr/

Xie, Y., Allaire, J. J., & Grolemund, G. (2018). R markdown: The definitive guide. Boca Raton, Florida: Chapman; Hall/CRC. Retrieved from https://bookdown.org/yihui/rmarkdown

Xu, H. (2007). The effects of self-construal and perceived control on privacy concerns. ICIS 2007 Proceedings, 1–14.

Zeileis, A., Hornik, K., & Murrell, P. (2009). Escaping RGBland: Selecting colors for statistical graphics. Computational Statistics & Data Analysis, 53(9), 3259–3270. https://doi.org/10.1016/j.csda.2008.11.033

Zhu, H. (2019). KableExtra: Construct complex table with 'kable' and pipe syntax. Retrieved from https://CRAN.R-project.org/package=kableExtra

Zuiderveen Borgesius, F. (2015). Behavioural sciences and the regulation of privacy on the internet. OxfordHart. Retrieved from https://dare.uva.nl/search?identifier=b0052c52-9815-4782-b4b0-b1cabb3624d0

Zuiderveen Borgesius, F. (2015a). Improving privacy protection in the area of behavioural targeting. Kluwer Law International. Retrieved from https://hdl.handle.net/11245/1.434236

Zuiderveen Borgesius, F., Hoboken, J. van, Fahy, R., Irion, K., Rozendaal, M., (2017). An assessment of the Commission's proposal on privacy and electronic communications: Study. European Parliament, Committee on Civil Liberties Retrieved from http://www.europarl.europa.eu/RegData/etudes/STUD/2017/583152/IPOL_STU(2017)583152_EN.pdf

Zuiderveen Borgesius, F., Kruikemeier, S., Boerman, S. C., & Helberger, N. (2017a). Tracking walls, take-it-or-leave-it choices, the GDPR, and the ePrivacy Regulation. European Data Protection Law Review, 3. https://doi.org/10.21552/edpl/2017/3/9

## APPENDIX A

R language and packages

We used R (Version 4.0.2; R Core Team, 2020) and the R-packages brms (Version 2.13.5; Bürkner, 2017, 2018), colorspace (Version 1.4.1; Zeileis, Hornik, & Murrell, 2009; Stauffer, Mayr, Dabernig, & Zeileis, 2009), dplyr (Version 1.0.1; Wickham et al., 2020), ggplot2 (Version 3.3.2; Wickham, 2016), gridExtra (Version 2.3; Auguie, 2017), here (Version 0.1; Müller, 2017), kableExtra (Version 1.1.0; Zhu, 2019), knitr (Version 1.29; Xie, 2015), papaja (Version 0.1.0.9942; Aust & Barth, 2020), pastecs (Version 1.3.21; Grosjean & Ibanez, 2018), plyr (Version 1.8.6; Wickham et al., 2020; Wickham, 2011), psych (Version 2.0.7; Revelle, 2019), Rcpp (Version 1.0.5; Eddelbuettel & François, 2011; Eddelbuettel & Balamuta, 2017), rmarkdown (Version 2.3; Xie, Allaire, & Grolemund, 2018), stringr (Version 1.4.0; Wickham, 2019), tidybayes (Version 2.1.1; Kay, 2020), tidyr (Version 1.1.1; Wickham & Henry, 2020), tokenizers (Version 0.2.1; Mullen, Benoit, Keyes, Selivanov, & Arnold, 2018), and VIM (Version 6.0.0; Kowarik & Templ, 2016) for all analyses and reporting.

# APPENDIX B

Consent requests

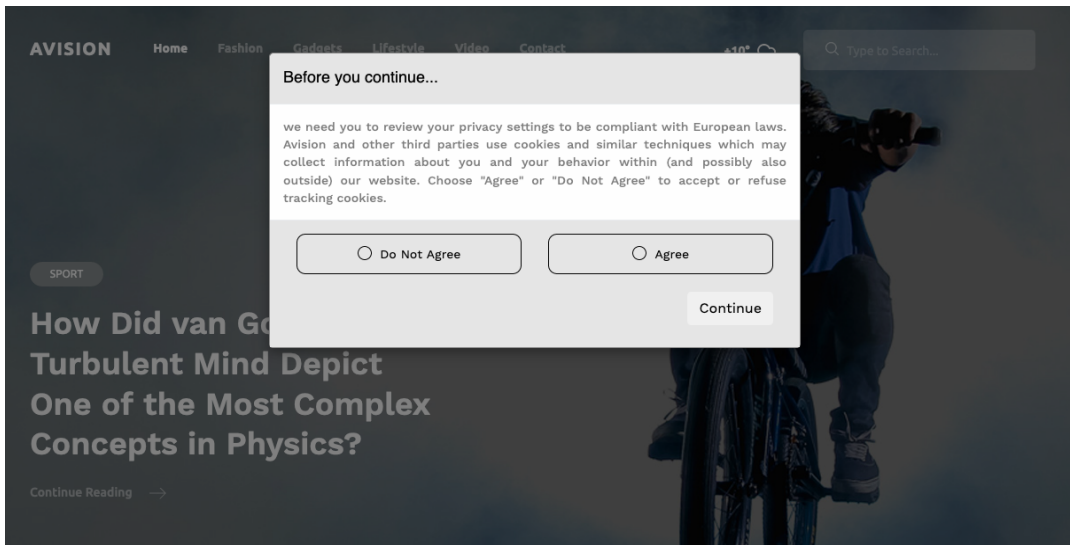

*Figure B1. Example Condition 1. Baseline. Website: Avision*

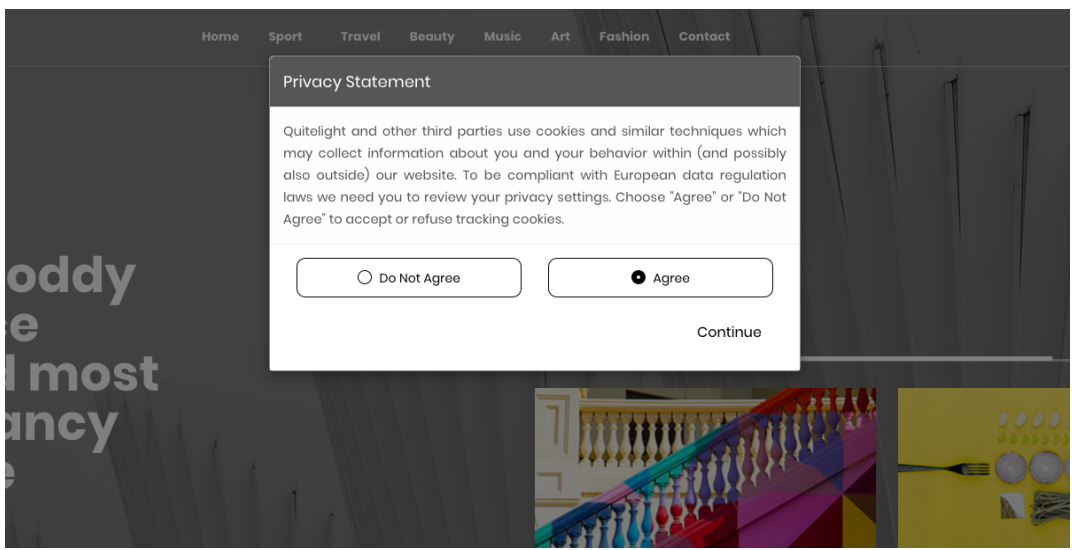

*Figure B2. Example Condition 2. Default. Website: Quitelight*

# APPENDIX C

Example consent request text from condition 1, news website Avision:

*"we need you to review your privacy settings to be compliant with European laws. Avision and other third parties use cookies and similar techniques which may collect information about you and your behavior within (and possibly also outside) our website. Choose 'Agree' or 'Do Not Agree' to accept or refuse tracking cookies."*